\documentclass[manuscript]{aastex}

\def\h0 {$h_0$=70 km s$^{-1}$ Mpc$^{-1}$}
\newcommand{\rej}{RE J1034+396}

\newcommand{\be}{\begin{equation}}
\newcommand{\ee}{\end{equation}}
\newcommand{\ce}{\ifmmode {\cal E} \else ${\cal E}$\ \fi}
\newcommand{\kms}{\ifmmode {\rm km\ s}^{-1} \else km s$^{-1}$\ \fi}
\newcommand{\ergs}{\ifmmode {\rm erg\ s}^{-1} \else erg s$^{-1}$\ \fi}
\newcommand{\tes}{\ifmmode \tau_{\rm es} \else $\tau_{\rm es}$\ \fi}
\newcommand{\tk}{\ifmmode \tau_{\rm K} \else $\tau_{\rm K}$\ \fi}
\newcommand{\vfwhm}{\ifmmode V_{\mbox{\tiny FWHM}} \else
            $V_{\mbox{\tiny FWHM}}$\fi}
\newcommand{\msun}{\ifmmode M_{\odot} \else $M_{\odot}$\ \fi}
\newcommand{\afe}{\ifmmode {\mathcal A_{\rm Fe}} \else${\mathcal A_{\rm Fe}}$\ \fi}

\newcommand{\lb}{\ifmmode L_{\rm Bol} \else $L_{\rm Bol}$\ \fi}
\newcommand{\ledd}{\ifmmode L_{\rm Edd} \else $L_{\rm Edd}$\ \fi}
\newcommand{\lx}{\ifmmode L_{\rm 2-10keV} \else  $L_{\rm 2-10keV}$\ \fi}
\newcommand{\hb}{\ifmmode H\beta \else H$\beta$\ \fi}
\newcommand{\mbh}{\ifmmode M_{\rm BH}  \else $M_{\rm BH}$\ \fi}
\newcommand{\lv}{\ifmmode \lambda L_{\lambda}(5100\AA) \else $\lambda L_{\lambda}(5100\AA)$\ \fi}

\def\ariel5{{\it Ariel 5}}

\def\xmm{{\it XMM-Newton}}
\def\chandra{{\it Chandra}}

\def\heao1{{\it HEAO~1}}

\def\rxte{{\it RXTE}}

\def\rxte{{\it RXTE}}

\shorttitle{CORRELATION BETWEEN BH MASS AND XVA}
\shortauthors{Zhou et al.}

\begin{document}
\title{Calibrating the correlation between black hole
mass and X-ray variability amplitude: X-ray only black hole mass
estimates for active galactic nuclei and ultra-luminous X-ray
sources }


\author{Xin-Lin Zhou\altaffilmark{}}
\affil{Key Laboratory of Optical Astronomy, National Astronomical
Observatories, Chinese Academy of Sciences, Beijing 100012, China}
 \affil{Department of
Physics and Tsinghua Center for Astrophysics, Tsinghua University,
Beijing 100084, China}

\author{Shuang-Nan Zhang\altaffilmark{}}
\affil{Key Laboratory of Particle Astrophysics, Institute of High Energy Physics, Chinese Academy of Sciences,,
Beijing 100049, China} 
\affil{ Physics Department, University of Alabama in Huntsville, Huntsville, AL 35899,
USA}

\author{Ding-Xiong Wang \altaffilmark{}}
\affil{School of Physics, Huazhong University of Science and Technology,
Wuhan 430074, China}

\author{Ling Zhu \altaffilmark{}}
 \affil{Department of
Physics and Tsinghua Center for Astrophysics, Tsinghua University,
Beijing 100084, China}

\email{zhouxl@nao.cas.cn}

\begin{abstract}

A calibration is made for the correlation between the X-ray
Variability Amplitude (XVA) and black hole (BH) mass. The
correlation for 21 reverberation-mapped Active Galactic Nuclei
(AGNs) appears very tight, with an intrinsic dispersion of 0.20 dex.
The intrinsic dispersion of 0.27 dex can be obtained if BH masses
are estimated from the stellar velocity dispersions. We further test
the uncertainties of mass estimates from XVAs for objects which have
been observed multiple times with good enough data quality. The
results show that the XVAs derived from multiple observations change
by a factor of 3. This means that BH mass uncertainty from a single
observation is slightly worse than either reverberation-mapping or
stellar velocity dispersion measurements; however BH mass estimates
with X-ray data only can be more accurate if the mean XVA value from
more observations is used. With this calibrated relation, the BH
mass and accretion rate may be determined for a large sample of AGNs
with the planned International X-ray Observatory mission.

Proper interpretation of the first AGN X-ray quasi-periodic
oscillation (QPO), seen in the Seyfert galaxy RE J1034+396, depends
on its BH mass, which is not currently known very well. Applying
this relation, the BH mass of RE J1034+396 is found to be
$4^{+3}_{-2} \times 10^6$ $M_{\odot}$. The high end of the mass
range follows the relationship between the 2$f_0$ frequencies of
high-frequency QPO and the BH masses derived from the Galactic X-ray
binaries.

We also calculate the high-frequency constant $C= 2.37~ M_\odot$
Hz$^{-1}$ from 21 reverberation-mapped AGNs. As suggested by
Gierli\'nski et al., $M_{\rm BH}=C/C_{\rm M}$, where $C_{\rm M}$ is
the high-frequency variability derived from XVA. Given the similar
shape of power-law dominated X-ray spectra in ultra-luminous X-ray
sources (ULXs) and AGNs, this can be applied to BH mass estimates of
ULXs. We discuss the observed QPO frequencies and BH mass estimates
in the ULXs M82 X-1 and NGC 5408 X-1 and favor ULXs as intermediate
mass BH systems.

\end{abstract}


\section{Introduction}
Active galactic nuclei (AGNs) are scaled-up versions of the Galactic
X-ray binaries (XRBs), as recently evidenced by the physical link
between the characteristic timescales and black hole (BH) masses
corrected by the accretion rates (McHardy et al. 2006). The
characteristic timescales are denoted by the break frequencies in
the power spectral densities (PSDs). This suggests a similar broken
power-law shape of PSDs from XRBs to AGN (e.g., Edelson \& Nandra
1999;
 Uttley, McHardy \& Papadakis 2002; Markowitz et al. 2003). The power spectra of XRBs also show
multiple types of quasi-periodic oscillations (QPOs) other than the
broadband noise. The first AGN QPO discovered in the Seyfert galaxy
\rej ~(Gierli\'nski et al. 2008a) stresses the similarities between
the XRBs and AGNs, but its proper interpretation is hammered by its
unknown BH mass. \rej ~is an extreme Narrow-Line Seyfert 1 galaxy
(NLS1), with the BH mass remaining largely uncertain (e.g., Bian \&
Zhao 2004). The BH mass determination is then a key issue since
there are many controversies on the NLS1 mass estimates (e.g.,
 Komossa \& Xu 2007; Marconi et al. 2008; Zhu et al. 2009).

We estimate the BH mass of \rej~ from the X-ray variability
amplitude (XVA), $\sigma^2_{\rm rms}$ (Nandra et al. 1997; Turner et
al. 1999). A major difficulty in estimating BH masses from their XVA
is the large scattering in the $M_{\rm BH}$$-$$\sigma^2_{\rm rms}$
relation (Lu \& Yu 2001; Bian \& Zhao 2003; Papadakis 2004; O'Neill
et al. 2005; Zhou et al. 2007; Miniutti et al. 2009; Nikolajuk et
al. 2009). Note that the previous works were based on the BH masses
derived from heterogeneous methods. This may cause additional
scattering in the relation, since different methods of BH mass
estimates may have quite different systematic errors. We therefore
attempt to make a calibration of $M_{\rm BH}$$-$$\sigma^2_{\rm rms}$
correlation based on only the reverberation-mapped BH masses; we
further check this calibration result with BH masses determined from
the dispersion velocity measures of these AGNs' host galaxies. We
also derive the high frequency constant as defined by Gierli\'nski
et al. (2008b). This is useful for the BH mass estimates of
ultra-luminous X-ray sources (ULXs) in nearby galaxies, since both
ULXs and AGNs have characteristic power-law dominated X-ray spectra.

This work is organized as follows. In Section 2, we re-analysis the
\xmm~observations of \rej~ and extract the light curves. In Section
3, we make a calibration of the correlation between the BH masses
and XVAs and derive the high frequency constant. In Section 4, we
discuss the uncertainties of mass estimates from XVA. In Section 5
we estimate BH mass for \rej ~and discuss the implication
 of the AGN QPO. We find \rej~ follows the relationship between the high-frequency
 QPO frequencies and BH masses derived from XRBs. Throughout
this work, we use the cosmological parameters of \h0 ,
$\Omega_{m}=0.27$, and $\Omega_{\Lambda}=0.73$.

\section{\xmm~ observations of \rej}
\xmm ~observed \rej~ on  2002 May 01 for 15.7 ks and
 on 2007 May 31 for 93 ks.
 We screen the 93 ks data with the SAS v7.0 software.
The X-ray events corresponding to patterns $0-4$ for the EPIC PN
data are selected.
 We extract the source light curves from a $45 ''$ circle of the detected
source position for PN, with the background being taken from six source-free regions
 with the same size. The presence of background flares and data gaps in the final 7 ks
are checked and removed.

Figure 1 shows the extracted $0.3-10$ keV PN light curve (128 s
bin$^{-1}$), which is similar in shape to that in Gierli\'nski et
al. (2008a). The light curve shows a periodic oscillation since
$t_{0}=25$ ks, with the frequency $f= 1/3730$~s $\approx2.7\times
10^{-4}$ Hz. Gierli\'nski et al. (2008a) have adopted the method
proposed by Vaughan (2005) to quantify the statistical significance
of periodicities against an underlying continuum which involves
dividing the periodogram by the best-fitting power law and using the
known distribution of the periodogram ordinates to estimate the
likelihood of observing a given peak. This method can be used to
test the significance of candidate periodicities superposed on a red
noise spectrum which has a power-law shape. The result indicates
that the QPO is at $\sim$5.6 $\sigma$ significance level.

\section{Calibrating $M_{\rm BH}$$-$$\sigma^2_{\rm rms}$ relation }

XVA is the variance of a light curve normalized by its mean squared after correcting for experimental noise. For
a light curve segment with $N$ bins (Nandra et al. 1997; Turner et al. 1999),
\begin{equation}
\label{sigma2}
 \sigma^{2}_{\rm rms}=\frac{1}{N\mu^{2}}\sum_{i=1}^{N} [(X_{i}-\mu)^2-\sigma_{i}^{2}],
\end{equation}
where $X_{i}$ and $\sigma_{i}$ are the counting rates and uncertainties in each bin.
 $\mu$ is the arithmetic mean of the counting rates.

Following O'Neill et al. (2005), the errors of  $\sigma^2_{\rm rms}$, which depend on the measurement
uncertainties and the stochastic nature of the source, can be expressed as,
 \begin{equation}
\label{sigma2er}
\Delta_{\mathrm{tot}}(\sigma^{2}_{\rm rms})      =      \sqrt{     \left(      \frac{
\sigma_{\mathrm{frac}}  \sigma^{2}_{\rm rms}}{\sqrt{N_{\mathrm{seg}}}}  \right)^{2} +
[\Delta_{\mathrm{boot}}(\sigma^{2}_{\rm rms})]^{2} },
\end{equation}
where $N_{\mathrm{seg}}$ is the number of available light-curve segments, $\sigma_{\mathrm{frac}}$ is a
fractional standard deviation, $\sigma_{\mathrm{frac}}=0.74$ for log$M_{\mathrm{BH}}>$ 6.54 and
$\sigma_{\mathrm{frac}}=0.48$ for log$M_{\mathrm{BH}}<$ 6.54. $\Delta_{\mathrm{boot}}(\sigma^{2}_{\rm rms})$ is
the bootstrap uncertainty which comes from
 the bootstrap simulation accounting for the measurement uncertainties.

XVA is calculated from a self-consistent analysis of an AGN's X-ray observation with a duration of at least 40
ks.
 The $2-10$ keV light curves are subdivided into segments with a
duration of 39.936 ks (binned at 256 s). The mean XVA is used if
there are multiple segments. We take the measured XVA listed in
O'Neill et al. (2005)
 if available.

\subsection{Reverberation-mapped sample}

Homogeneous BH masses are required for studying and calibrating the
$M_{\rm BH}$-$\sigma^2_{\rm rms}$ relation. BH masses based on the
reverberation-mapping (RM) method for 35 AGN (Peterson et al. 2004)
and the small Seyfert galaxy NGC 4395 (Peterson et al. 2005) are
currently available.
 Here we use the reverberation-mapped BH masses for the
calibration of the $M_{\rm BH}$$-$$\sigma^2_{\rm rms}$ relation. We
list in Table 1 the 21 reverberation-mapped AGN with at least one
segment of X-ray data longer than 40 ks.

In Figure 2, we show the correlation between XVA and the RM BH
masses. We assume that there is a linear correlation between the
$M_{\rm BH}$ and XVA, $y=\alpha +\beta x$, where $y=\log(M_{\rm
BH}/M_\odot)$, $x=\log(\sigma^2_{\rm rms})$, with the measurement
errors of $\epsilon_{x_i}$ for $x_i$ and $\epsilon_{y_i}$ for $y_i$.
It was shown that the Nukers' estimate
 is an unbiased slope estimator for
the linear regression (Tremaine et al. 2002). The Nukers' estimate is based on minimizing,

\be \chi^2\equiv
\sum_{i=1}^N{(y_i-\alpha - \beta x_i)^2\over \epsilon_{yi}^2+\beta^2\epsilon_{xi}^2}.
 \label{eq:chisq} \ee

We apply the Nukers' estimate to derive the parameter of
$\alpha=4.85\pm0.20$ and $\beta = -1.05 \pm 0.08$ for the 21 RM
AGNs. The minimum $\chi^2$ per degree of freedom is 1.6, indicating
that either the uncertainties in the BH masses are underestimated or
there is an intrinsic dispersion for BH mass estimates.

To account for the intrinsic dispersion in the $M_{\rm
BH}$$-$$\sigma^2_{\rm rms}$ relation, we replace $\epsilon_{yi}$ by
$(\epsilon_{yi}^2+\epsilon_0^2)^{1/2}$, where $\epsilon_0$
represents the intrinsic dispersion; $\epsilon_0$ is adjusted so
that the value of $\chi^2$ per degree of freedom is unity. This
procedure is preferable if the individual error estimates
$\epsilon_{y_i}$ are reliable (Tremaine et al. 2002).
 Adding an intrinsic dispersion
of 0.18 dex decreases the value of $\chi^2$ per degree of freedom to
unity, and gives the best-fit results, $\alpha=4.97\pm0.26$, and
$~\beta=-1.00\pm0.10$. Then,

\be \label{eq:fitresults} M_{\rm BH}= 10^{4.97\pm0.26}(\sigma^2_{\rm
rms})^{-1.00\pm0.10} M_{\odot}.
\ee

Note that the errors of the RM BH mass for IC4329A and NGC 4593 are
quite large (Peterson et al. 2004). We exclude these two objects to
ensure reliable error estimates of BH masses. The intrinsic
dispersion of the $M_{\rm BH}$-$\sigma^2_{\rm rms}$ relation is 0.20
dex for the rest of 19 AGNs. We also plot the objects with the BH
masses estimated from the empirical virial relation; these objects
(open circles in Figure 2) generally follow the best-fit correlation
well.


\subsection{Sample with stellar velocity dispersions}
The key result we obtained is that the intrinsic dispersion of the
$M_{\rm BH}$$-$$\sigma^2_{\rm rms}$ relation is quite small, no
larger than the uncertainties of RM BH masses of $0.3-0.4$ dex
(Peterson et al. 2004). We further check this result by using the BH
masses estimated from the stellar velocity dispersions, since AGN
hold the same relation between the BH masses and the stellar
velocity dispersions (hereafter $M_{\rm BH}$$-$$\sigma_*$ relation)
as the normal galaxies (Gebhardt et al. 2000; Ferrarese et al.
2001). In Table 2, we list 21 AGNs with the stellar velocity
dispersion measurements; only four objects are different from that
listed in Table 1.

Figure 3 shows the relation of XVA against the BH mass. The BH mass
is estimated from the relation $M_{\rm BH}=1.349\times10^8
(\sigma_*/200)^{4.02} M_{\odot}$ (Tremaine et al. 2002). However,
there is evidence for large scattering at the low BH mass end in the
$M_{\rm BH}$-$\sigma_*$ relation (Greene \& Ho 2006). We therefore
use the RM mass of NGC 4395 (Peterson et al. 2005), and estimate the
$M_{\rm BH}$ of MCG-6-30-15 from the empirical virial relation using
the \hb line (Zhou \& Wang 2005). It is difficult to determine the
actual errors of mass estimates in Table 2.  Tremaine et al. (2002)
claimed that the stellar velocity dispersion shows a Gaussian
measurement error of 0.06 dex. Assuming no intrinsic dispersion in
the $M_{\rm BH}$$-$$\sigma_*$ relation, the error of mass estimates
from the stellar velocity dispersion should be $4.02\times0.06
\approx 0.24$ dex. Therefore, we add the error of $\pm0.24$ dex for
the BH masses in Table 2.

We then apply Nukers' estimate to derive the parameter of
$\alpha=4.91\pm0.24$ and $\beta = -1.04 \pm 0.09$. The minimum
$\chi^2$ per degree of freedom is 1.8, indicating that either the
uncertainties in $M_{\rm BH}$  are underestimated or there is an
intrinsic dispersion for $M_{\rm BH}$ estimates. We add an intrinsic
dispersion of 0.27 dex to decrease the value of $\chi^2$ per degree
of freedom to unity, and give the best-fit results,
$\alpha=5.15\pm0.29$, and $~\beta=-1.01\pm0.11$. Then,

\be \label{eq:sresults} M_{\rm BH}= 10^{5.15\pm0.29}(\sigma^2_{\rm
rms})^{-1.01\pm0.11} M_{\odot}. \ee

The results are similar to that from the RM sample. This is a
natural result since the RM method has been calibrated in agreement
with the $M_{\rm BH}$-$\sigma_*$ relation (Peterson et al. 2004). If
the errors of $M_{\rm BH}$ used here are underestimated, the
intrinsic dispersion of the $M_{\rm BH}$$-$$\sigma^2_{\rm rms}$
relation should be smaller. Since it is difficult to determine the
actual errors of $M_{\rm BH}$ estimates used here, we tend to use
the calibrated relation with the RM sample, i.e., Equation (4).

\subsection{Weak dependence on Eddington ratio}
$\sigma^2_{\rm rms}$ may also depend on the second variable,
Eddington ratio, $\dot{m}_{\rm E}$ (e.g., Papadakis 2004). Following
McHardy et al. (2006), we hypothesize that log$M_{\rm BH}= A + B{\rm
log}\sigma^2_{\rm rms}+C{\rm log}L_{\rm bol}$. The fit is good with
$A=-(0.23\pm0.04)$, $B=-(0.87\pm0.12)$, and $C=0.12\pm0.03$. Using
$\dot{m}_{\rm E}= L_{\rm bol}/L_{\rm Edd}$, then $\sigma^2_{\rm rms}
\propto \dot{m}_{\rm E}^{0.14} / M_{\rm BH}$, indicating that the BH
accretion rates only weakly affect the mass estimates.

We further explore the dependence of $\sigma^2_{\rm rms}$ on the X-ray luminosity
(e.g., Barr \& Mushotzky 1986; Liu \& Zhang 2008),
  FWHM of the broad components of the
optical H$\beta$ lines (e.g., Leighly 1999), and the X-ray photon
index (e.g., O'Neill et al. 2005). Results of partial correlation
analysis show that the correlation between $\sigma^2_{\rm rms}$ and
$M_{\rm BH}$ are the main correlation after removing the effects of
other dependence (Table 3).

\subsection{Amplitude of High-frequency Variability}
Gierli\'nski et al. (2008b) have argued that the XVA at high frequencies can be used as an estimator of a BH's
mass. Their hypothesis is that there is a universal power spectral shape for BHs at high frequencies (e.g.,
Lawrence \& Papadakis 1993; Hayashida et al. 1998) above the PSD break frequency, $f_b$, as illustrated in Fig.
4. This universal form is roughly a power law with an index $\alpha=2$, then $P(f)=C_M (f/f_r)^{-2}$, where
$f_r$ is an arbitrary frequency chosen to be $f_r = 1$ Hz.

Then XVA is calculated between frequencies $f_1$ and $f_2$, with $f_b \leq f_1 < f_2$,
\begin{equation}\sigma_{\rm rms}^2 = \int_{f_1}^{f_2} P(f) df
= C_M f_r \left( {f_r \over f_1} - {f_r \over f_2}
\right).\label{eq:sigma_nxs}\end{equation}

 The assumption is that $C_{\rm M}$ is inversely
proportional to the BH mass, $M_{\rm BH}=C/C_{\rm M}$, where $C$ is
the high frequency constant. Gierli\'nski et al. (2008b) calculated
the $C$ values from the Galactic XRB Cyg X-1. Different $C$ values
 can be obtained if the data corresponding to the different spectral state of Cyg X-1 are used;
$C=1.24~M_\odot$ Hz$^{-1}$ for the low hard state and $C=2.77
~M_\odot$ Hz$^{-1}$ for the soft state.

We calculate $C$ from the 21 AGNs with the RM masses. We find
$C=2.37~M_\odot$ Hz$^{-1}$, close to the value for the soft state of
Cyg X-1. This means that $C$ is not sensitive to $M_{\rm BH}$ since
the BH mass crosses over 8 orders of magnitudes from XRBs to AGNs.
This also agrees with the fact that the amplitude of PSDs only
slightly changes from XRBs to AGNs (see Fig. 18 in McHardy et al.
2004). Thus,, $C$ can be reasonably assumed to be constant for AGN
mass estimates.

We emphasize that the $C$ value from AGNs can be also applied for
the BH mass estimates of ULXs in nearby galaxies, because the X-ray
spectral properties of ULXs are similar to that of AGNs. Both ULXs
and AGNs show power-law dominated X-ray spectra, together with a
soft X-ray excess. Also, both ULXs and AGNs lack the
disk-emission-dominated state as frequently observed in the Galactic
XRBs.

\subsection{$f_1$ test and origin of intrinsic dispersion}
In Figure 5 we show the dependence of XVA on $f_1$. If $f_1 \ll
f_2$, $\sigma_{\rm rms}^2\approx C_M f_{r}^{2}/f_1$, where $f_1$
corresponds to the lowest frequency of data sampled.
 That means, XVA is inversely proportional to $f_1$ used. We calculate XVAs
 from a segment of 40 ks \xmm~ data with a different $f_1$. XVAs generally decrease with the increasing $f_1$, but
 only a few AGNs follow the slope of $-1$. The measured deviation from $\sigma_{\rm rms}^2 \propto 1/f_1$ may
 be due to either a red-noise component in the light curve (Vaughan et al. 2003), or generally
 the assumption of $\alpha=2, f_b \leq
 f_1 \ll f_2$ is not always right for the sample. It was also known that the PSD of NGC
4051 shows a break frequency of $8\times 10^{-4}$ Hz (McHardy et al.
2004), which is higher than the $f_1$ used (for this work,
$f_1=1/39.936~{\rm ks} = 2.5 \times 10^{-5}$ Hz). However, PSDs of
most of AGNs in the sample are still unknown. These may contribute
to the intrinsic dispersion of the $M_{\rm BH}$$-$$\sigma^2_{\rm
rms}$ relation.

\section{Errors of $M_{\rm BH}$ estimates from XVA}
The intrinsic dispersion of the $M_{\rm BH}$$-$$\sigma^2_{\rm rms}$
relation derived above is at a level of $0.2-0.3$ dex, no larger
than the uncertainties of BH masses used here. Can the mass
estimates from XVA be more accurate than those estimates from
 the empirical virial
relation or from the stellar velocity dispersion? In this section we
discuss the errors of the mass estimates from XVA.

X-ray PSDs of AGN are not stationary on the timescales probed by
typical \xmm~ or \chandra~ observations.
 The red-noise nature of X-ray
variability of AGN has been established (e.g., Lawrence et al. 1987;
McHardy \& Czerny 1987). For the red-noise process, XVA depends on
the shape of PSD, and the time resolution and duration of the light
curve. XVA may change from one part of the light curve to the next,
even when the variability is produced by a constant PSD. Thus,
different XVA values can be obtained if different data are used.
Therefore, we attempt to estimate the uncertainties of mass
estimates by looking at objects which have been observed multiple
times with good enough data quality to make independent $M_{\rm BH}$
estimates.

Figure 6 shows the distribution of XVA for the NLS1 galaxy
MCG-6-30-15 and 1H0707-495. MCG-6-30-15 was observed using \xmm~
starting on 2000 July 11 and ending on 2001 August 4, with a total
exposure
  time of $\sim$480 ks. 1H0707-495 was observed using \xmm~ starting on 2000 October 21
  and ending on 2008 February 6, with a total exposure time of $\sim$800 ks.
  We analyze all these data and filter the light
  curves to remove any background flares. We then subdivide the light
  curves into the 39.936 ks segments binned at 256 s. The XVA derived from
  each data segment shows the variability by a factor of 3 ($\sim$0.5 dex). This
  is roughly consistent with the error estimates by Equation (2).

Gierli\'nski et al. (2008b) have analyzed  \rxte~ Proportional
Counter Array observations of seven Galactic XRBs. The
high-frequency variability derived from XVA changes by a factor of
$2-3$, even when the data corresponding to different spectral state
are used. Therefore the error of mass estimates from a single data
segment is slightly larger than the RM method of $0.3-0.4$ dex,
despite of the small intrinsic dispersion of the $M_{\rm
BH}$$-$$\sigma^2_{\rm rms}$ relation. However, assuming XVA randomly
scatters around the true value for the PSD of the process (Vaughan
et al. 2003), the mean XVA of many data segments may reduce the
error and thus the mass estimates can be more accurate than with the
RM or stellar velocity dispersion method.

We would also like to point out that the RM AGNs are a subset of BHs
which are skewed toward low mass, high Eddington fraction of AGNs.
These are probably the bulk of the AGNs for which reliable XVA can
be computed with the present instrumentation. It would be not safe
to attempt to apply this method to the objects like M31* or Sgr A*
with the same calibration, since they both have extremely low
accretion rates. The XVA can be a function of the photon energy.
Thus the choice of spectral bandpass affects the XVA estimates.
Nevertheless we do not address this issue in the
 current work since all the results presented here are based on the
 $2-10$ keV band. However, if an X-ray instrument with a softer bandpass were used, or if one wanted to apply this
 method to higher redshift AGNs where both time dilation and redshifting of the X-rays were relevant,
 it would be necessary to develop an understanding of what would change.
 This will be investigated in a future work.

\section{BH mass and QPO frequency}
\label{qpo}

\subsection{BH Mass of \rej}
 The X-ray variability of \rej~ increases with the photon energy (Middleton et
al. 2009). We extract the $2-10$ keV light curve for the XVA
calculation. The mean XVA of the two segments of light curves of
\rej~ is found to be $0.0234\pm0.0081$, corresponding to a BH mass
of $4^{+3}_{-2} \times 10^6$ \msun (See Figure 2).

BH masses in AGN are usually determined from two methods: RM method
and the $M_{\rm BH}$$-$$\sigma_*$ relation. We apply the empirical
virial relation based on the RM method firstly and then compare the
result from the stellar velocity dispersion, in order to estimate
the BH mass in RE J1034+396. The gas in the broad line region (BLR)
are thought to be virialized. The virial BH mass can then be
calculated from

\be M_{\rm BH} = R_{\rm BLR} f^{2}{\rm FWHM^{2}_{H\beta}}/G , \ee
 where FWHM$_{\rm H\beta}$
is the FWHM of the whole \hb. For an isotropic velocity
distribution, as generally assumed, $f = \sqrt{3} /2$. $R_{\rm BLR}$
is obtained from the RM experiment or the $L_{\rm 5100}$$-$$R_{\rm
BLR}$ relation calibrated with the RM method. The BH mass in RE
J1034+396 calculated from Equation (7) is $6.3 \times 10^{5}$
$M_{\odot}$, when FWHM$_{\rm H\beta} = 701$ km s$^{-1}$ is used
(Bian \& Zhao 2004). However, recently it has been found that for
NLS1s, the broad \hb~line can be usually decomposed with a double
Gaussian model, i.e., a very broad gaussian component (VBGC) and an
intermediate gaussian component (IMGC) (Mullaney \& Ward 2008; Zhu
et al. 2009). It is suggested that the BH mass should be calculated
with FWHM of the VBGC (Zhu et al. 2009). Then, the virial BH mass
should be corrected as \be M_{\rm BHb} = R_{\rm VBLR} f^2{\rm
FWHM^2_{H\beta}(FWHMb/FWHM_{\hb})}^2/G , \ee
  where
FWHMb is FWHM$_{\rm \hb}$ of the VBGC and $R_{\rm VBLR}$ is the
radius of the very broad line region (VBLR). $R_{\rm BLR}$, is used
as $R_{\rm VBLR}$, because RM usually measures the size of the VBLR
(Zhu et al. 2009). Its \hb line can be well decomposed into three
components as shown in Mason et al (1996): a narrow line component,
VBGC and IMGC. Bian \& Huang (2009) have recently analyzed the SDSS
spectrum of RE J1034+396, by properly modeling the spectrum of its
host galaxy. They found that the FWHM of the VBGC is $1690 \pm 296$
km s$^{-1}$. We use Equation (8) to correct the BH mass, $M_{\rm
BHb} = (1690/701)^2\times 6.3 \times 10^5$ \msun $= 3.7\times 10^6
~\msun$. Considering the possible intrinsic scattering in the virial
mass determination, the mass range of \rej~ should be $2-7 \times
10^6~\msun$. This mass is in excellent agreement with the result
from XVA, also roughly consistent with the range of
$1-4\times10^6~\msun$, determined from the stellar velocity
dispersion of the host galaxy of RE J1034+396 (Bian \& Huang 2009).

\subsection{QPO in RE J1034+396}
Figure 7 shows the scale of accretion flows between the BH mass and
the QPO frequency across six orders of magnitudes. For a typical
high-frequency (HF) QPO, $f_0=931(M_{\rm BH}/\msun)$ Hz (Remillard
\& McClintock 2006), where $f_0$ is the fundamental frequency. This
relation is derived from three Galactic XRBs, XTE J1550$-$564, GRO
J1655$-$40 and GRS 1915$+$105, which display a pair of HFQPOs with
3:2 frequency ratios. The frequencies of XRBs are plotted for the
stronger QPO that represent $2 \times f_{0}$. If the QPO in \rej~ is
a typical high-frequency QPO with $f = 2 f_0$ or $f = 3f_0$, it
would correspond to a BH mass of $6.9\times 10^6$ \msun or
$1.0\times10^7 $ \msun, respectively. The high end of mass range of
\rej~ favors this AGN QPO as a HFQPO with $f = 2f_0$. Middleton \&
Done (2009) suggested that the QPO is the AGN analog of the 67 Hz
QPO in GRS 1915+105, and not the HFQPOs seen in 3:2 ratios. However,
that one scales the QPO frequency for
  the ($4\times10^6/14$) mass ratio between RE J1043+396 and GRS 1915+105,
  one gets 76 Hz $-$ within $\sim10\%$ of 67 Hz, and well within the error bars on the BH mass for RE
J1043+396.

HFQPOs with the harmonic pairs of frequencies in a 3:2 ratio were
often suggested to arise from some type of resonance mechanism
(e.g., Abramowicz \& Kluzniak 2001; Wang et al. 2007), where the
frequencies are correlated with the BH mass and spin. The BH spin
only changes the resonance orbits of accreting matter by a factor of
a few. These may be the underlined physical reasons for the precise
scale of accretion flows between the HFQPO frequency and BH mass.
Rezzolla et al. (2003) proposed a model which requires that the two
frequencies being oscillation modes of the same torus, rather than
requires a resonance between the two QPO frequencies. The advantage
of the Rezzolla et al. model for explaining what is seen here is
that the higher frequency QPO is not seen, and the strengths of the
different modes can vary independently. An alternative explanation
for not seeing the higher frequency QPO might be that the PSD has
not been made in the $10-30$ keV range where that QPO is usually
seen in the Galactic XRBs, so this is not conclusive evidence in
favor of the Rezzolla model over the resonance models. The QPOs
previously seen are stellar mass BHs, which form through stellar
evolution and grow insubstantially. They might all have quite
similar spins. AGN, which can grow by both mergers and accretion,
could have a different range of BH spin. Hence even if the QPO is
due to the same mechanism as what produces the 3:2 frequency ratio
QPOs in XRBs, a very realistic possibility is that the BH in RE
J1043+396 has a mass closer to the most likely mass of $4\times10^6$
$M_{\odot}$ from the XVA analysis, but has a lower spin than the
stellar mass BHs.

\subsection{QPOs in ULXs}

There are QPOs observed in the ULX M82 X-1 (Strohmayer \& Mushotzky
2003; Feng \& Kaaret 2007) and NGC 5408 X-1 (Strohmayer et al.
2007), but the BH masses of these two ULXs were estimated ranging
from 25 to 10$^5$ \msun (e.g., Dewangan et al. 2006; Wu \& Gu 2008,
etc). The non-standard accretion disk scenario (e.g., Soria et al.
2007) can explain the low soft X-ray temperature of M82 X-1 without
invoking the intermediate mass BH. However, by assuming the M82 X-1
similar to the low-luminosity hard state of Galactic XRBs, Yuan et
al. (2007) derived a mass of $9-50\times10^4$ \msun for M82 X-1 by
fitting the \chandra~ X-ray data together with the constraints of
radio and infrared upper limits using the advection-dominated
accretion flow model. Lower BH mass ($10^4$ $M_{\odot}$) can be
obtained if M82 X-1 corresponds to the high-luminosity hard state.
Assuming the QPOs in M82 X-1 and NGC 5408 X-1 are low-frequency
QPOs, Casella et al. (2008) placed the constraints on the BH masses
of $100-1300$ \msun. Obviously, other possible QPO identifications
will yield to different mass estimates. From the correlation between
the spectral slope and the QPO frequency (known to exist in Galactic
XRBs, see e.g. Vignarca et al. 2003), Fiorito \& Titarchuk (2004)
derived a mass of the order of $10^3$ \msun for M82 X-1, although no
such correlation was found in ULXs. Recently, Strohmayer \&
Mushotzky (2009) found that the X-ray spectral and timing properties
of NGC 5408 X-1 are quite analogous to the Galactic XRBs in the
``very high'' or ``SPL'' state. They also derived the high-frequency
variability, $C_{\rm M}$ (Gierli\'nski et al. 2008b) for mass
estimates. If we adopt the high-frequency constant $C=2.37~M_\odot$
Hz$^{-1}$ from this work, their $C_{\rm M}$ values correspond to
mass ranges of $3200-4000$ \msun and $8400-10800$ \msun for the two
observations. This is not far from the scale between the HFQPO
frequencies and the BH masses.

\section{Conclusions}
Calibration of the correlation between the black hole mass and the
XVA shows that this correlation is quite tight, with an intrinsic
dispersion of around 0.2 dex. The intrinsic dispersion may be caused
by the uncertain shape of power spectral densities and the
stochastic nature of the red-noise
 light curves. The independent mass estimates from the X-ray variability derived from
multiple observations show the uncertainties by a factor of 3
($\sim$0.5 dex). This is slightly larger than the errors of
reverberation mapping or the stellar velocity dispersion
measurements. However, the BH mass estimates can be more accurate if
using the mean XVA derived from multiple observations.
 BH accretion rates only weakly affect the mass estimates.
We find the high frequency
 constant $C=2.37 M_\odot$ Hz$^{-1}$ from the reverberation-mapped
 AGN.  Given the similar shape of
power-law dominated X-ray spectra in Ultra-Luminous X-ray sources
and AGN, this can be applied to mass estimates of ULXs in nearby
galaxies. The BH mass of RE J1034+396 is found to be $4^{+3}_{-2}
\times 10^6$ \msun from the XVA. The high end of mass range of RE
J1034+396  follows the relation between the high-frequency QPO
frequencies and the BH masses derived from the Galactic X-ray
binaries, suggesting the precise scaling of accretion flows among
them. We also discuss the observed QPO frequencies and BH mass
estimates in the ULX M82 X-1 and NGC 5408 X-1 and favor the
intermediate mass BH systems.

 We also find that X-ray variability as the mass estimator is independent of the luminosity, the
 distance and the viewing angle of the source; where luminosity is calculated by assuming the isotropy of emission.
 This is not true in many cases.
 Jets and outflows are common
  features and thus the beaming effect may play a role. The viewing angle may also change the inferred
  luminosity (Zhang 2005) considerably, since the X-ray emission from an accretion disk is generally not
  isotropic. Nevertheless, XVA provides quite reliable BH mass measurements without knowing
 the ``true" luminosity of the source.
  Conventionally the BH mass in an AGN is often estimated from the measures of its broad line's width
  which may depend on the
 viewing angle of the source, unless the broad line region is completely spherically symmetric.
For example, a narrower line profile can be obtained if it is seen
from the pole-on angle and the BLR is flattened with a substantial
Keplerian motion component. The radiation pressure may also affect
the scale of the BLRs in AGNs. Apparently, XVA is independent of
these effects.
   Therefore, the method of estimating BH's masses with XVAs may be a good alternative for NLS1 galaxies and ULXs,
   whose BH mass estimates have remained largely uncertain so far. Long and deep X-ray survey observations may allow
   the BH masses and thus accretion rates determined for a large sample of AGN, with, for example, the planned
   International X-ray Observatory.

\acknowledgments We are very grateful to an anonymous referee for
many useful comments and suggestions, which allowed us to improve
the manuscript substantially. We thank useful discussions with Bian,
W.-H., Feng, H., Laor, A., Li, T. P., Liu, Y., Lu, Y., Soria, R.,
Wang, J.-M., Ye, Y.-C., Zhang, Y.-H. and Zhao, Y.-H. We also thank
Tremaine, S. to answer our questions on the fitting algorithm. This
work was supported, in part, by China postdoctoral science
foundation,
 Directional Research Project of the CAS under project No. KJCX2-YW-T03, the National Natural Science
 Foundation of China under grant Nos. 10821061, 10733010, 10725313, and 973 Program of China under grant
2009CB824800.




\clearpage

\begin{figure}
\includegraphics[width=11 cm, angle=270]{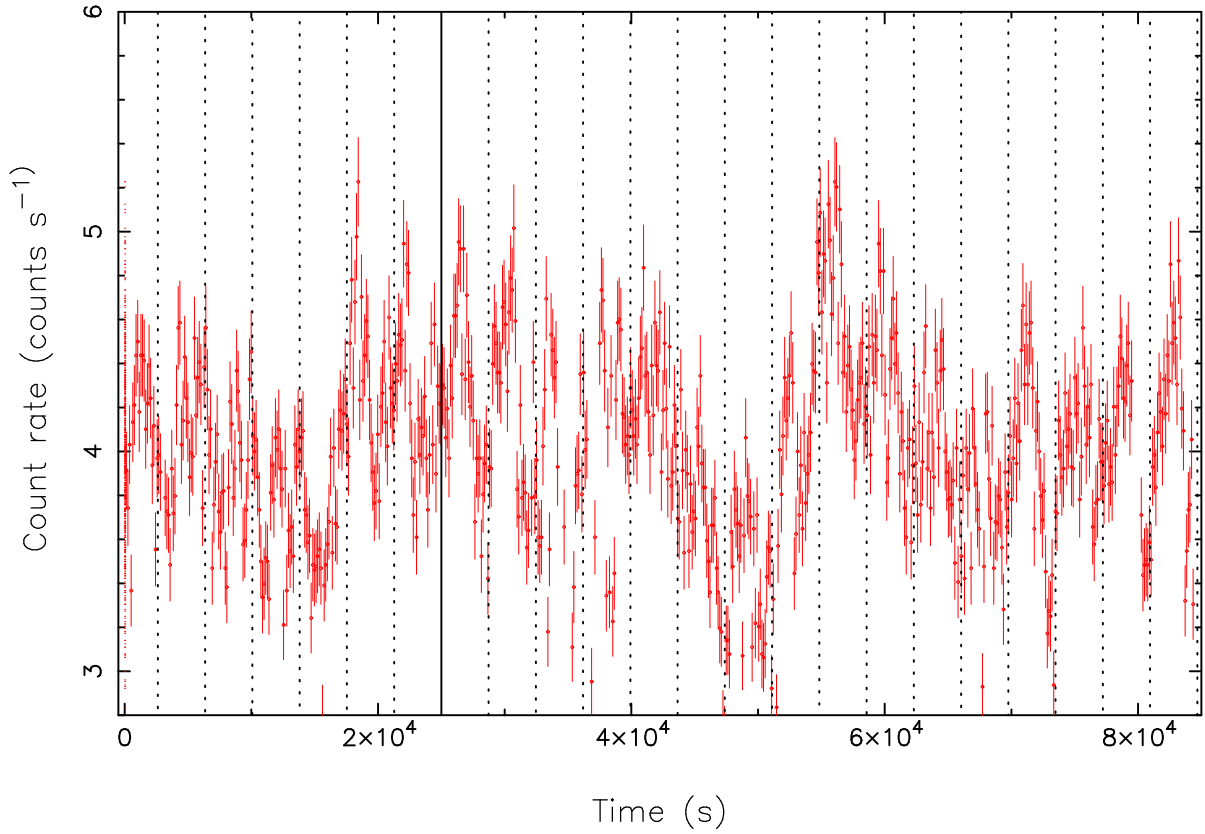}\label{fig1}
\caption  {The source light curve extracted from the \xmm~PN camera in the $0.3-10$ keV band binned at 128 s for
\rej. It is similar in shape to that in Gierli\'nski et al. (2008a). It shows a periodic
oscillation of $T=3730$ s since $t_{0}=25$ ks.}
\end{figure}


\begin{figure}
\includegraphics[width=9 cm, angle=270]{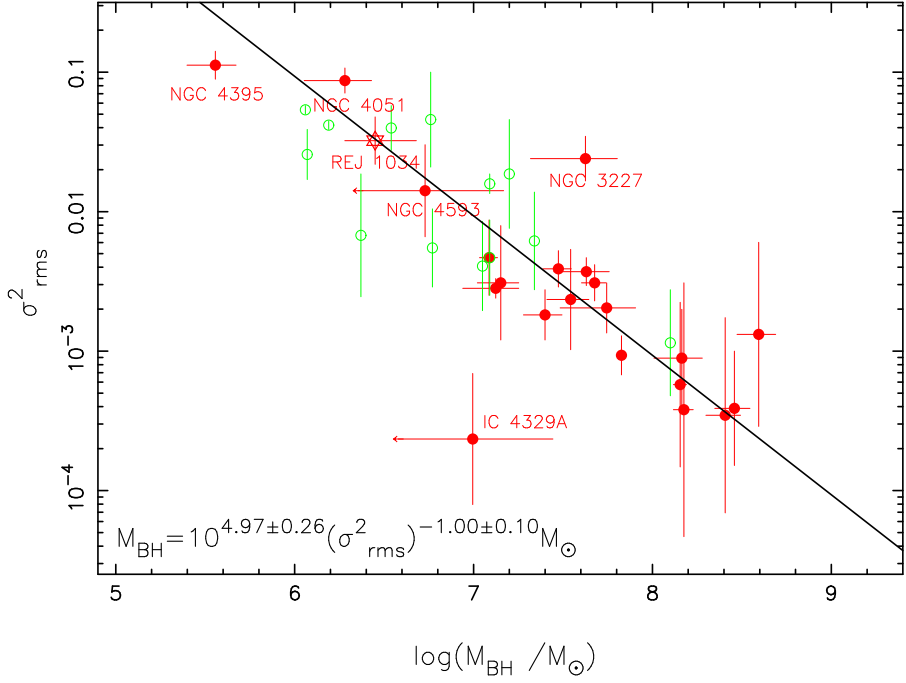}\label{fig2}
\caption  {Correlation between the X-ray variability amplitude and the BH mass. The filled points denote the 21
objects with the BH masses measured from the reverberation-mapping method.  The best-fit linear relation given
by the Nukers' estimate for the 21 AGN is also shown in the plot. The intrinsic dispersion of this fit is 0.2
dex. Most of objects show negligible scattering compared with the linear fit. The star denotes RE J1034+396.  We
also plot the objects with the mass estimated from the empirical virial relation; these objects (open circles)
generally well follow the linear fitting relation. }
\end{figure}

\begin{figure}
\includegraphics[width=9 cm, angle=270]{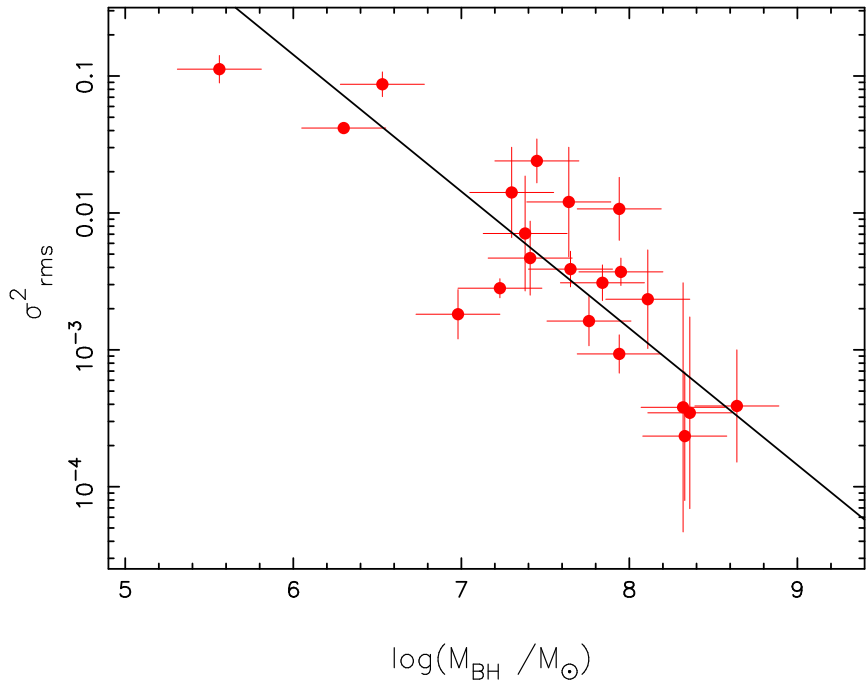}\label{fig3}
\caption  {Correlation between the X-ray variability amplitude and the
 BH mass. The BH masses are mainly estimated from the stellar velocity dispersion measurements,
 with an assumed error of 0.24 dex.  See the text for the details. The
 Nuker's estimates give the similar results to that in Fig. 2, with an intrinsic dispersion of 0.27 dex. }
\end{figure}

\begin{figure}
\includegraphics[width=9 cm, angle=270]{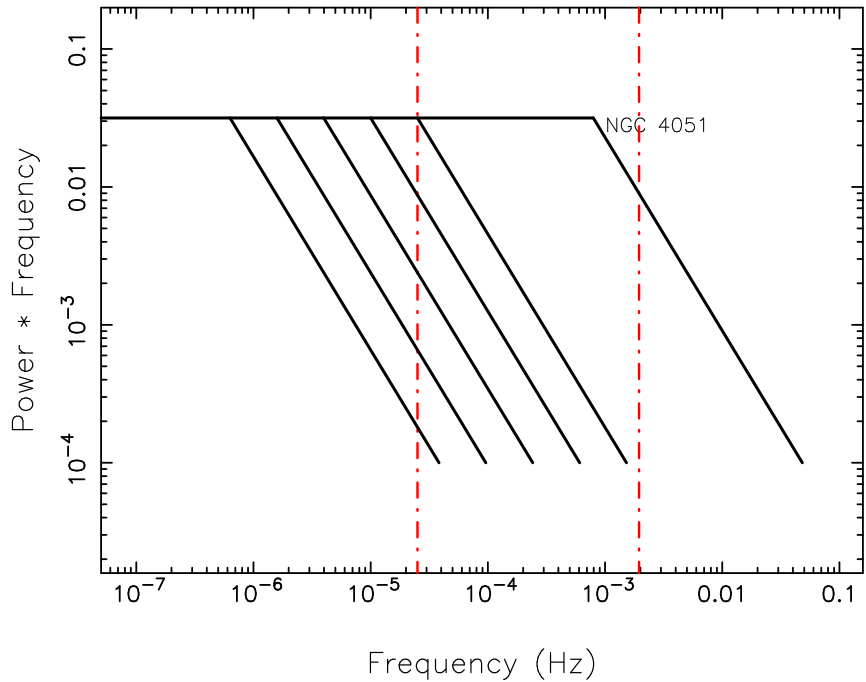}\label{fig:psd}
\caption {The assumed power-law shape of the power spectral densities (PSD) at high frequency. The two
dot-dashed lines denote the low frequency $f_1$ and the high frequency $f_2$ sampled. For this work,
$f_1=1/39.936~{\rm ks} = 2.5 \times 10^{-5}$ Hz, and $f_2=1/2 \times 256~{\rm s} = 2.0 \times 10^{-3}$ Hz.
 To ensure the estimation of
the integral over the power law PSD, the frequency break should be lower than $f_1$. However, NGC 4051 has a
higher frequency break of $8 \times 10^{-4}$ Hz. This may cause the deviation of NGC 4051 in the $M_{\rm
BH}-\sigma^2_{\rm rms}$ correlation.}
\end{figure}

\begin{figure}
\includegraphics[width=9 cm, angle=270]{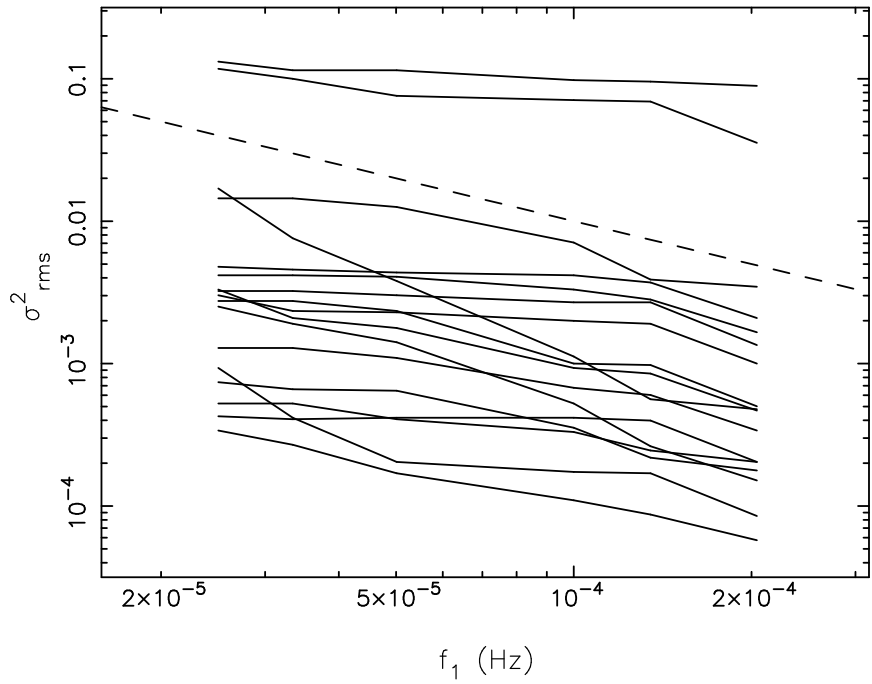}\label{fig:f1test}
\caption { Dependence of X-ray variability amplitude
 on $f_1$ from a segment of 40 ks \xmm~ data.
 $\sigma^2_{\rm rms}$ generally decreases with the increasing $f_1$, but
 only a few AGN follow the slope of $-1$ (dashed line).}
\end{figure}

\begin{figure}
\includegraphics[width=9 cm, angle=270]{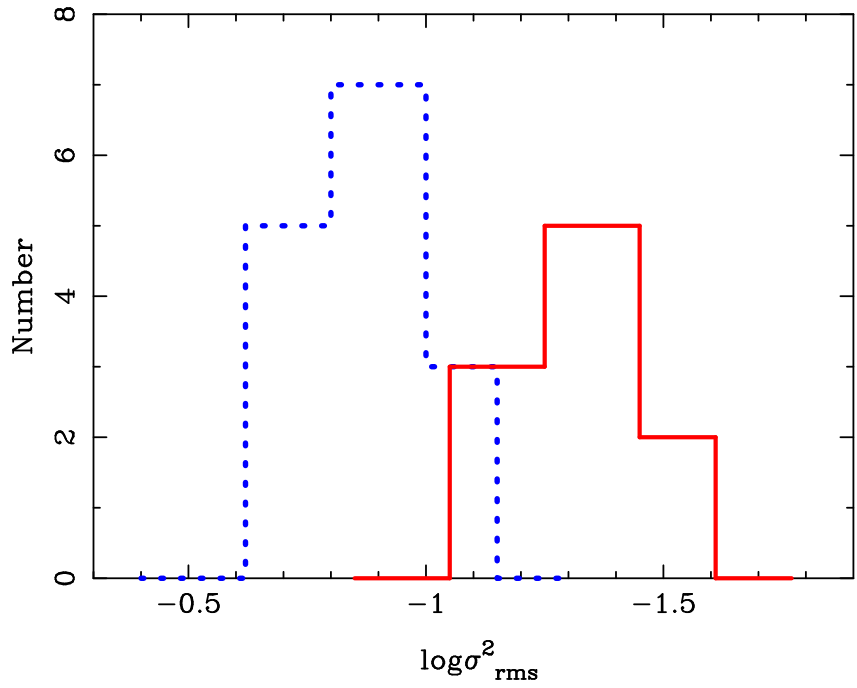}\label{fig:nrms}
\caption { Distribution of X-ray variability amplitude derived from multiple \xmm~observations for the NLS1
galaxy MCG-6-30-15 (solid line) and 1H0707-495 (dotted line).  }
\end{figure}

\begin{figure}
\includegraphics[width=9 cm, angle=270]{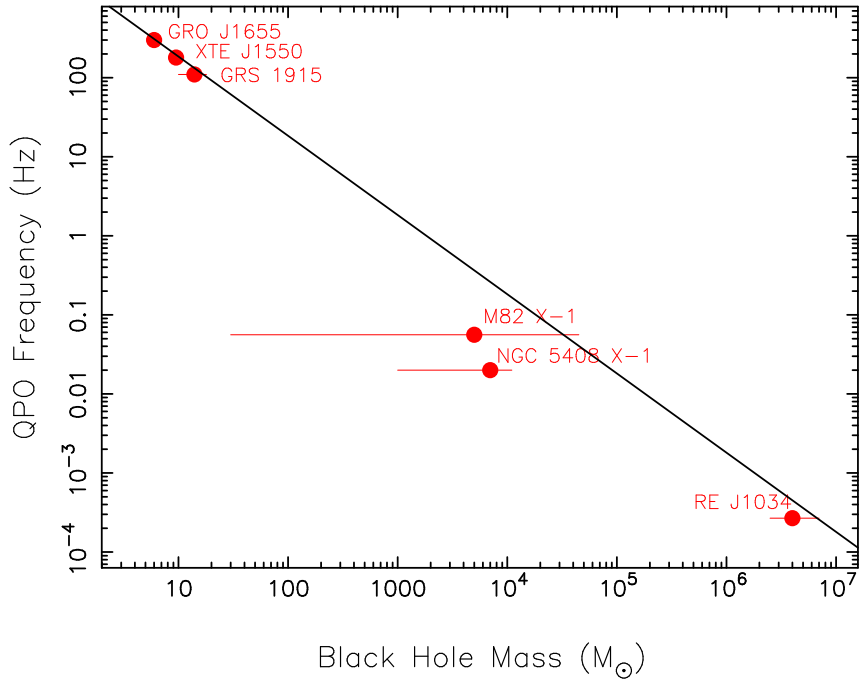}\label{figqpo}
\caption  {Relationship between the quasi-periodic oscillation
(QPO) frequency and the BH mass. Three X-ray binaries, XTE
J1550$-$564, GRO J1655$-$40 and GRS 1915+105 display a pair of
high-frequency QPOs with a 3:2 frequency ratio. The frequencies of X-ray binaries
are plotted for the stronger QPO that represent $2 \times f_{0}$.
The line denotes the relation, $f(Hz)=1862 (M_{\rm BH}
/M_{\odot})^{-1}$ (Remillard \& McClintock 2006). The observed QPO frequency in the ULX M82 X-1
and NGC 5408 X-1 is also plotted. See the text for details.}
\end{figure}

\begin{table*}[t1]
\begin{center}
\footnotesize \centerline{\sc Table 1. Sample of AGN with
reverberation-based mass } \vglue 0.1cm
\begin{tabular}{llllcrc}\hline \hline
Name        &$z$    &$M_{\rm BH}$ & log$\sigma_{\rm rms}^{2}$ & Num.
Seg.     & log$L_{\rm bol}$ & log$\dot m_{\rm E}$  \\
(1)         & (2)    & (3)  & (4)  &(5)&  (6)   & (7)
\\\hline
Mrk 335    &0.02578 &  14.2$\pm$3.7         & $-2.51\pm$0.41    & 1  &  $44.81$  & $-0.44$  \\
PG 0026+129 & 0.14200     &  393 $\pm$ 96         & $-2.88\pm$0.66    & 3  &  $45.97$ & $-0.72$  \\
Fairall 9  &0.04702 &  255$\pm$56 & $-3.46\pm0.70$    & 6    & 45.20 & $-1.31$ \\
Mrk 590    &0.02638 &  47.5$\pm$7.4 & $-2.51\pm0.13$ & 2 & 44.76  & $-1.02$ \\
3C 120     &0.03301 &  55.5$^{+31.4}_{-22.5}$   & $-2.79\pm 0.18$ &  $ 4 $ & 45.12 & $-0.72$  \\
Ark 120    &0.03230 &  $150\pm19 $ & $-3.42\pm$0.91    & 2   & 45.18 & $-1.10$ \\
Mrk 110   &0.03529 &  $25.1\pm6.1$ & $-2.74\pm0.18$ & 1 & 44.67 & $-0.83$  \\
NGC 3227   &0.00386 &  $42.2\pm21.4$ & $-1.62\pm$0.16  & 4 &  43.33 & $-2.40$ \\
NGC 3516   &0.00884 &  $42.7\pm14.6$ & $-2.43\pm$0.10 & 18 &  43.83 &  $-1.89$ \\
NGC 3783   &0.00973 &  $29.8\pm5.4$  & $-2.41\pm0.13$   & 8 & 44.21 & $-1.36$ \\
NGC 4051   &0.00234 &  $1.91\pm 0.78$       & $-1.06\pm$0.09  & 6 &42.88 & $-1.50$ \\
NGC 4151   &0.00332 &  $13.3\pm 4.6$       & $-2.55\pm$0.07   & 29 &43.83 & $-1.39$\\
PG 1211+143&0.08090 &  $146 \pm 44 $     & $-3.05\pm0.35$ &3 & 45.70 & $-0.56$  \\
NGC 4395   &0.00106 & $0.36\pm0.11$         & $-0.95\pm$0.10 & 6 & 41.42 & $-2.24$  \\
NGC 4593   &0.00900 & $5.36^{+9.37}_{-6.95}$         & $-1.85\pm$0.33  & 1 & 44.03 & $-0.79$ \\
IC 4329A   &0.01605 & $9.90^{+17.88}_{-11.88}$ & $-3.63\pm$0.47  & 6 & 44.27 & $-0.83$ \\
Mrk 279    &0.03045 & $34.9\pm9.2$ & $-2.63\pm$0.36  & 1  & 44.83 & $-0.81$  \\
NGC 5548   &0.01718 & $49.4\pm7.7$   & $-3.03\pm0.14$  & 16  &  44.46 & $-1.47$\\
3C 390.3   &0.05610 & $287\pm64$         &  $-3.41\pm0.41$ &2 &44.90 & $-1.66$  \\
Mrk 509    &0.03440 & $143\pm12$        &  $-3.24\pm0.59$ & 2  & 45.23 & $-1.03 $ \\
NGC 7469   &0.01632 & $12.2\pm 1.4$          &  $-2.33\pm0.27$ &  2
& 44.67  & $-0.49$
\\\hline
\end{tabular}
{\baselineskip 8pt \indent\\
(1) Object name. (2) Redshift. (3) Black hole mass in units of
$10^{6}M_{\odot}$, taken from Peterson et al. (2004) and Peterson et
al. (2005). (4) Log of the mean X-ray variability amplitude. (5)
Number of usable light-curve segments (39.936 ks per segment). (6)
Log of bolometric luminosity. (7) Log of the Eddington ratio, $\dot
m_{\rm E}=L_{\rm bol}/L_{\rm Edd}$.}
\end{center}
\end{table*}
\normalsize

\begin{table*}[t2]
\begin{center}
\footnotesize \centerline{\sc Table 2. Sample of AGN with stellar
velocity dispersions} \vglue 0.1cm
\begin{tabular}{llccllc}\hline \hline
Name        &$z$    &log($M_{\rm BH}/M_{\odot})$ & $\sigma_*$ & Ref.  &
log$\sigma_{\rm rms}^{2}$ & Num.
Seg.      \\
(1)         & (2)    & (3)  & (4)  &(5)&  (6) & (7)
\\\hline
Fairall 9  &0.04702 &  $8.36$ & 228 & 4  & $-3.46\pm0.70$    & 6    \\
Mrk 590    &0.02638 &  $7.84$ & 169 &  2  & $-2.51\pm0.13$ & 2  \\
Mrk 1040   &0.01665 &  $7.64$ & 151 &  2  & $-1.92\pm0.40$ & 1    \\
3C 120     &0.03301 &  $7.76$ & 162  & 1  & $-2.79\pm 0.18$  &  4   \\
Ark 120    &0.03230 &  $8.32$ & 224  & 1 & $-3.42\pm0.91$    & 2    \\
Mrk 79     &0.02219 &  $7.38$ & 130  & 3 & $-2.15\pm0.42$  & 1  \\
Mrk 110   &0.03529 &   $6.99$  & 87 &  1   & $-2.74\pm0.18$ & 1  \\
NGC 3227   &0.00386 &  $7.45$   & 135 & 1 & $-1.62\pm$0.16  & 4 \\
NGC 3516   &0.00884 &  $7.95$   & 181 & 1 & $-2.43\pm$0.10 & 18  \\
NGC 3783   &0.00973 &  $7.65$   & 152  & 4  & $-2.41\pm0.13$   & 8 \\
NGC 4051   &0.00234 &  $6.53$ & 80  &  3     & $-1.06\pm$0.09  & 6  \\
NGC 4151   &0.00332 &  $7.23$ & 119  &  2  & $-2.55\pm$0.07   & 29  \\
NGC 4395   &0.00106 & $5.56$  & 30 & 1      & $-0.95\pm$0.10 & 6   \\
NGC 4593   &0.00900 & $7.30$    & 124  & 2     & $-1.85\pm$0.33  & 1  \\
MCG-6-30-15&0.00775 & $6.30$ &  159   & 4 & $-1.38\pm0.03$ & 48   \\
IC 4329A   &0.01605 & $8.34$  & 225 & 4  & $-3.63\pm$0.47  & 6  \\
NGC 5506   &0.00618 & $7.94$    & 180   &  4   & $-1.97\pm0.23$ & 2     \\
Mrk 279    &0.03045 & $8.11$& 197   & 1  &  $-2.63\pm$0.36  & 1  \\
NGC 5548   &0.01718 & $7.94$   & 180  & 3  & $-3.03\pm0.14$  & 16  \\
3C 390.3   &0.05610 & $8.64$&  268 &  1       &  $-3.41\pm0.41$ &2 \\
NGC 7469   &0.01632 & $7.41$   & 133 &1 & $-2.33\pm0.27$ &
2 \\\hline
\end{tabular}
{\baselineskip 8pt \indent\\
(1) Object name. (2) Redshift. (3) Black hole mass, mainly estimated
from the stellar velocity dispersion. See the text for the details.
(4) stellar velocity dispersion, in unit of km s$^{-1}$. (5)
references for stellar velocity dispersions, 1. Greene \& Ho (2006);
2. Nelson \& Whittle (1995); 3. Ferrarese et al. (2001); 4. Oliva et
al. (1999). (6) Log of the mean X-ray variability amplitude. (7)
Number of usable light-curve segments (39.936 ks per segment). }
\end{center}
\end{table*}
\normalsize

\begin{table*}[t3]
\begin{center}
\footnotesize \centerline{\sc Table 3. Partial correlation analysis}
\vglue 0.1cm
\begin{tabular}{lllllcrccr}\hline \hline
$L_{2-10 \rm keV}$, $M_{\rm BH}$      &  $r_{M\sigma}$    &$r_{L\sigma}$ & $r_{ML}$ & $r_{M\sigma,L}$  & $r_{L\sigma,M}$ \\
                  & $-0.85$ & $-0.69$ & $0.73$ & $-0.70$ & $-0.19$ \\\hline
FWHM\hb, $M_{\rm BH}$   &  $r_{M\sigma}$    &$r_{F\sigma}$ & $r_{MF}$ & $r_{M\sigma,F}$  & $r_{F\sigma,M}$ \\
                         &  $-0.85$          &  $-0.52$ &  $0.33$  & $-0.84$ &$-0.48$  \\\hline
$\Gamma_{2-10 \rm keV}$, $M_{\rm BH}$   &  $r_{M\sigma}$    &$r_{\Gamma\sigma}$ & $r_{M\Gamma}$ & $r_{M\sigma,\Gamma}$ & $r_{\Gamma\sigma,M}$\\
                                        &  $-0.85$          & $-0.20$ & ... & ... & ... \\\hline
\end{tabular}
\parbox{3.45in}
{\baselineskip 8pt \indent\\
 $r$ is the correlation coefficient. $M$, $L$, $F$ and $\Gamma$
denotes $M_{\rm BH}$, $L_{2-10 \rm keV}$, FWHM\hb ~and $\Gamma_{2-10 \rm
 keV}$, respectively.}
\end{center}
\end{table*}
\normalsize

\end{document}